\documentclass[useAMS,usedcolumn,usegraphicx,usenatbib]{mn2e}

\usepackage{amsmath}

\defcitealias{Josselin03}{JB03}

\title[CO$\bmath{^+}$ in PNe]
      {A search for CO$\bmath{^+}$ in planetary nebulae}
\author[T.~A.~Bell et al.]
{T.~A.~Bell,$^{1,2}$\thanks{E-mail: tab@submm.caltech.edu}
 W.~Whyatt,$^{1}$
 S.~Viti$^{1}$ and
 M.~P.~Redman$^{3}$\\
$^{1}$Department of Physics \& Astronomy, University College London, Gower Street, London WC1E 6BT\\
$^{2}$Department of Astronomy, California Institute of Technology, Pasadena, CA 91125, USA\\
$^{3}$Department of Physics, National University of Ireland Galway, Galway, Ireland}

\begin{document}

\date{Accepted 2007 August 27. Received 2007 August 8; in original form 2007 January 27}

\pagerange{\pageref{firstpage}--\pageref{lastpage}} \pubyear{2007}

\maketitle

\label{firstpage}

\begin{abstract}
We have carried out a systematic search for the molecular ion CO$^+$ in a sample of 8 protoplanetary and planetary nebulae in order to determine the origin of the unexpectedly strong HCO$^+$ emission previously detected in these sources. An understanding of the HCO$^+$ chemistry may provide direct clues to the physical and chemical evolution of planetary nebulae. We find that the integrated intensity of the CO$^+$ line may be correlated with that of HCO$^+$, suggesting that the reaction of CO$^+$ with molecular hydrogen may be an important formation route for HCO$^+$ in these planetary nebulae.
\end{abstract}



\section{Introduction}\label{Introduction}

Planetary nebulae (PNe) are commonly identified and studied by their optical emission arising in the ionized gas. However, CO and H$_2$ surveys have revealed that a large fraction of the gas in massive PNe is in fact in molecular form \citep[e.g.][]{Huggins96,Huggins05,Speck02,Pardo07}. Several studies \citep[e.g.][hereafter JB03]{Bachiller97,Josselin03} have shown that this molecular component may be dominant during a large part of the evolution of the nebulae and hence plays an important role in the mass distribution and the shaping of the PNe. The molecular gas in PNe will be mainly affected by the large radiation fields from the central stars: the dramatic effects on the chemical composition will change as the star moves from the asymptotic giant branch (AGB) into the protoplanetary nebula (PPN) phase and then into a PN \citep[see review by][]{Kwok93}.

Most chemical studies of the neutral gas in PNe have so far focused on comprehensive surveys of only a handful of PNe, all typically evolved, with the exception of CRL\,618 and NGC\,7027 \citep[e.g.][]{Cox92,Bachiller97,Pardo04,Pardo07}. However, \citetalias{Josselin03} recently reported high sensitivity observations of molecular lines in a sample of 7 young or intermediate-aged compact PNe, providing us with the best sample, so far, of PNe at \textit{different} evolutionary stages. Their conclusions can be summarized as follows: 1) some molecular species, such as HCO$^+$ and CN, are particularly abundant compared to envelopes around AGB stars or even to interstellar clouds; 2) a careful analysis of the chemical composition of a large enough sample \textit{can} give an indication of the evolutionary stage of the PPN/PN; 3) the survival of certain molecules can only occur in well protected dense clumps \citep[e.g.~the famous cometary knots in the Helix nebula and envelope,][]{Huggins02,Speck02}. The latter conclusion supports the theoretical findings of \citet{Redman03}, who showed that clumpiness in the slow wind from the central star may be responsible for the high neutral content of PNe, although other studies \citep[e.g.][]{Woods04} find that clumpiness in itself is not enough to protect the molecular component. Interestingly however, \citet{Redman03} and \citetalias{Josselin03} also agree in finding that HCO$^+$ and HCN are severely underpredicted by theoretical models and, in fact, HCO$^+$ is \textit{always} found to be remarkably high in PNe \citep[e.g.][]{Bachiller97}.

\begin{table*}
\begin{minipage}{130mm}
 \caption{Properties of the observed sample of (proto-)planetary nebulae.}
 \label{sources}
 \begin{tabular}{@{}l r@{\ }r@{\ }d{2} r@{\ }r@{\ }d{1} . r c c c}
  \hline
  Source & \multicolumn{3}{c}{RA(2000)} & \multicolumn{3}{c}{Dec.~(2000)} & \multicolumn{1}{c}{$D_{\rmn{K}}$} & \multicolumn{1}{c}{Size} & $V_{\rmn{LSR}}$ & $V_{\rmn{exp}}$ & References \\
  & \multicolumn{3}{c}{(hh mm ss)} & \multicolumn{3}{c}{\ ($\degr\ \quad\arcmin\ \quad\arcsec$)} & \multicolumn{1}{c}{(kpc)} & \multicolumn{1}{c}{(arcsec)} & ($\rmn{km\,s^{-1}}$) & ($\rmn{km\,s^{-1}}$) & \\
  \hline
  BV\,5--1 & 00 & 20 & 00.45 & $+62$ & 59 & 03.2 & 5.5 &      11\ \,\ \ & $-73$ & 10 & 1,2 \\
  K\,3--94 & 03 & 36 & 08.09 & $+60$ & 03 & 46.3 & 5.5 & $\la$10\ \,\ \ & $-69$ & 16 & 1,3 \\
  CRL\,618 & 04 & 42 & 53.67 & $+36$ & 06 & 53.2 & 1.8 &      14\ \,\ \ & $-22$ & 18 & 4,5 \\
  M\,1--13 & 07 & 21 & 14.95 & $-18$ & 08 & 36.9 & 2.0 & $\la$12\ \,\ \ & $+27$ & 18 & 1,6 \\
  M\,1--17 & 07 & 40 & 22.21 & $-11$ & 32 & 29.8 & 2.5 & $\la$12\ \,\ \ & $+28$ & 39 & 1,7 \\
  K\,3--24 & 19 & 12 & 05.82 & $+15$ & 09 & 04.5 & 3.5 & $\la$10\ \,\ \ & $+44$ & 24 & 1,3 \\
  NGC\,6781$^{\star}$ & 19 & 18 & 28.09 & $+06$ & 32 & 19.3 & 0.7 & 110\ \,\ \ & $+17$ & 22 & 4,8 \\
  IC\,5117 & 21 & 32 & 31.03 & $+44$ & 35 & 48.5 & 3.0 & $\la$12\ \,\ \ & $-11$ & 14 & 1,6 \\
  \hline
 \end{tabular}

 \medskip
 References: 1) \citet{Josselin03}; 2) \citet{Josselin01}; 3) \citet{Josselin00}; 4) \citet{Bachiller97}; 5) \citet{Sanchez04b}; 6) \citet{Huggins96}; 7) \citet{Bachiller91}; 8) \citet{Bachiller93}.\\
 $^{\star}$Since the angular size of this PN is significantly larger than the telescope beam, an offset position of $(+10\arcsec,+60\arcsec)$ was adopted, at which HCO$^+$ had previously been observed.
\end{minipage}
\end{table*}

Recent observational and theoretical work \citep[e.g. JB03;][]{Woods03} has shown that the post-AGB phase is a period of great chemical changes. The appearance, or enhancement, of HCO$^+$ may correspond to different evolutionary stages, depending on what causes its high abundance. For example, the standard path for HCO$^+$ formation,
\begin{equation}
 \rmn{H_{3}^{+} + CO \to HCO^{+} + H_{2}},
\end{equation}
is generally excluded for PNe because it requires a high survival rate of H$_3^+$, which is instead expected to be very short-lived due to its rapid dissociative recombination in the presence of a UV radiation field. Conversely, harder radiation fields due to, for example, strongly ionizing X-ray emission from the central star work to maintain high abundances of H$_3^+$ by increasing its production rate. If this is the cause of the high abundance of HCO$^+$ then the enhancement should occur at an early stage, just after the ionization of the nebula begins at the end of the PPN stage \citep{Kwok93}. An additional formation route for HCO$^+$ is through
the reaction of CO$^+$ with H$_2$,
\begin{equation}
 \rmn{CO^{+} + H_{2} \to HCO^{+} + H}.
\end{equation}
The reactive molecular ion CO$^+$ can be formed by charge transfer between H$^+$ and CO or by direct ionization of CO by means of the propagation of an ionizing front into molecular gas of high CO abundance \citep*{Latter93}. Thus, there are two formation routes for CO$^+$, one of which only becomes effective once the object is evolved enough for significant ionization of the nebula to take place (which occurs at late times). There are also two routes to form HCO$^+$, one that can only occur early in the evolution of the source and one that requires the presence of CO$^+$. Therefore, the varying abundances of the two molecules can potentially be used as an evolutionary diagnostic tool.

The obvious test to discriminate between the above scenarios is to perform a systematic search for CO$^+$ in PNe and correlate its emission with that of HCO$^+$, as, in fact, \citetalias{Josselin03} suggest. CO$^+$ has been clearly detected in one young PN, namely NGC\,7027 \citep{Latter93,Hasegawa01,Fuente03}. In this paper we present the results of such a search and show that it is likely that CO$^+$ is directly correlated with HCO$^+$, making these two ions good evolutionary tracers of PNe.


\section{Observations}

\subsection{PNe sample}\label{Sample}

We have selected a sample of 8 PPNe and PNe in which HCO$^+$ has previously been detected. Our sample is primarily based on that of \citetalias{Josselin03}, who detected the HCO$^+$(1--0) line in 7 young to intermediate-age compact PNe. In addition, we also include the PPN CRL\,618 and the evolved PN NGC\,6781 to increase the range of evolutionary stages in our sample. These two sources have also been previously detected in HCO$^+$(1--0) emission \citep{Bachiller97}. We do not include NGC\,7027, where CO$^+$ has already been detected \citep{Latter93,Hasegawa01,Fuente03}. Table~\ref{sources} lists the properties of the PNe in our sample; their common names (column 1), J2000 coordinates of their central positions (2, 3), kinematic distances (4), based on the simple approximation for the Galactic disc velocity structure of \citet{Burton74}, sizes (5) based on CO(2--1) maps (where available; upper limits are given when the nebulae are unresolved), radial and expansion velocities, as measured in the CO(2--1) line (6, 7) and the papers from which this information was obtained (8).

\subsection{Observations}\label{Observations}

The observations were obtained with the James Clarke Maxwell Telescope (JCMT), Mauna Kea, Hawaii on the nights of 2005 February 18--20 and 2005 October 2--10, with additional data collected on various dates in 2005. Heterodyne receivers RxA3i and RxB3 were used for the 200 and 300~GHz band observations, respectively. Transitions of CO$^+$ are described in terms of the quantum numbers $N$, the rigid body angular momentum excluding electron spin, and $J=N\pm1/2$, the total angular momentum. The CO$^+$ $N=2\to1$, $J=5/2\to3/2$ (236.0626~GHz) line was searched for in the sample of 8 PNe, whilst detection of the CO$^+$ $N=3\to2$, $J=7/2\to5/2$ (354.0142~GHz) line was only attempted in CRL\,618, the source displaying the strongest emission in the lower CO$^+$ line in our sample. The antenna half-power beamwidth (HPBW) was 20~arcsec at 236~GHz and 14~arcsec at 354~GHz. A frequency resolution of 625~kHz was initially used with an instantaneous bandwidth of 920~MHz, however the resolution was changed to 312.5~kHz and the bandwidth to 500~MHz for the data obtained in 2005 October. With the exception of NGC\,6781, all observations were carried out in beam-switching mode, with a secondary mirror throw of 60~arcsec in azimuth. Since NGC\,6781 has an angular size of 110~arcsec \citep{Bachiller93}, the secondary mirror throw was increased to 180~arcsec for this source to ensure that the nebula did not fall within the OFF position. Pointing and focus were checked every 60--90 minutes and the pointing uncertainty was found to be within 3~arcsec. The data were reduced in the standard manner using the \textsc{specx} software package and were converted to the main beam ($T_{\rmn{mb}}$) scale using main beam efficiencies determined by planetary observations made whilst the data were being acquired. During the data reduction process, several channels were binned and Hanning smoothing was applied to reduce the noise level. The channel binning was performed in the standard way, by averaging channels into bins 2--4 channels wide, starting from the first channel in the spectrum.


\section{Results}\label{Results}

\begin{figure*}
 \includegraphics[width=0.487\textwidth]{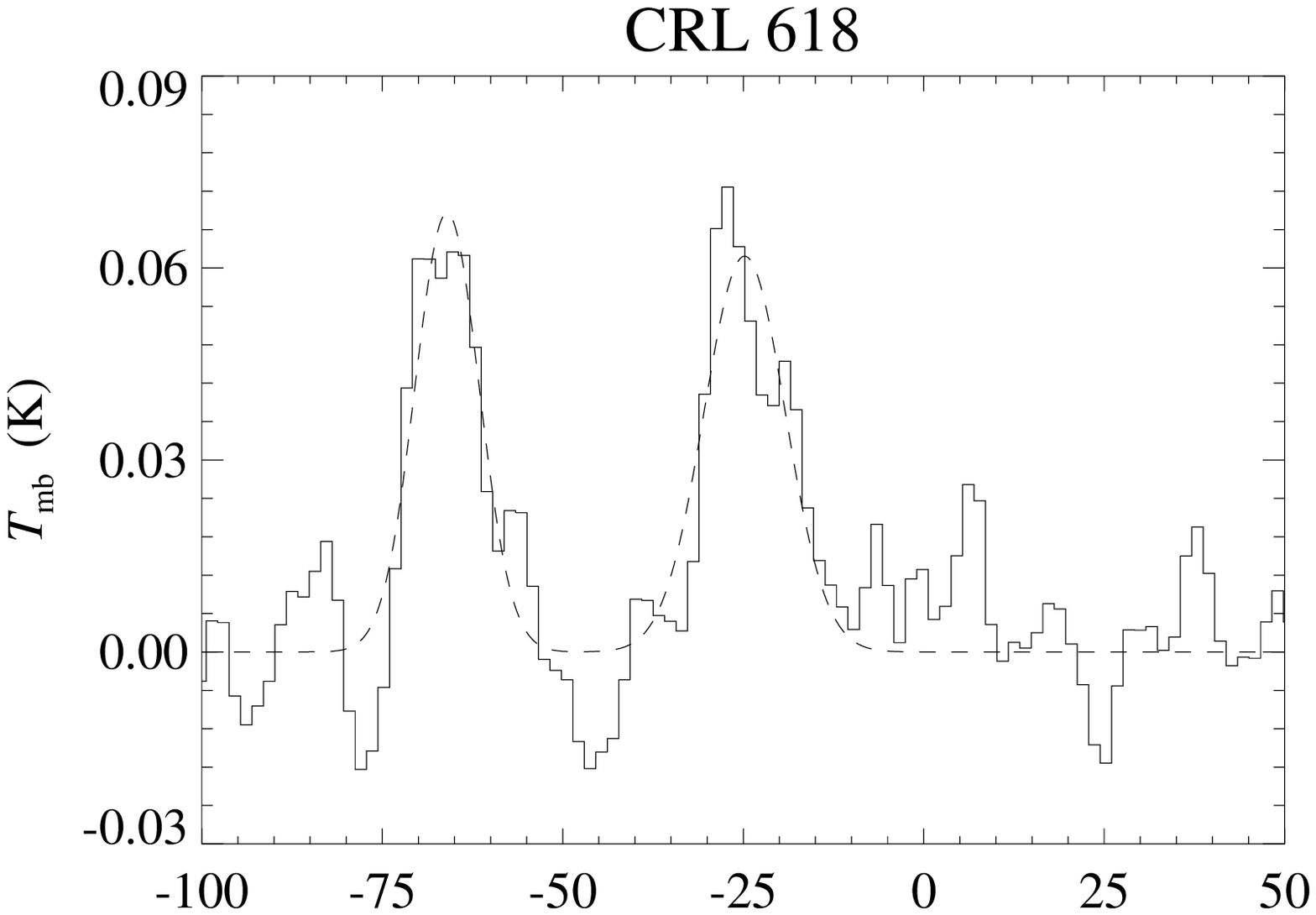}\includegraphics[width=0.463\textwidth]{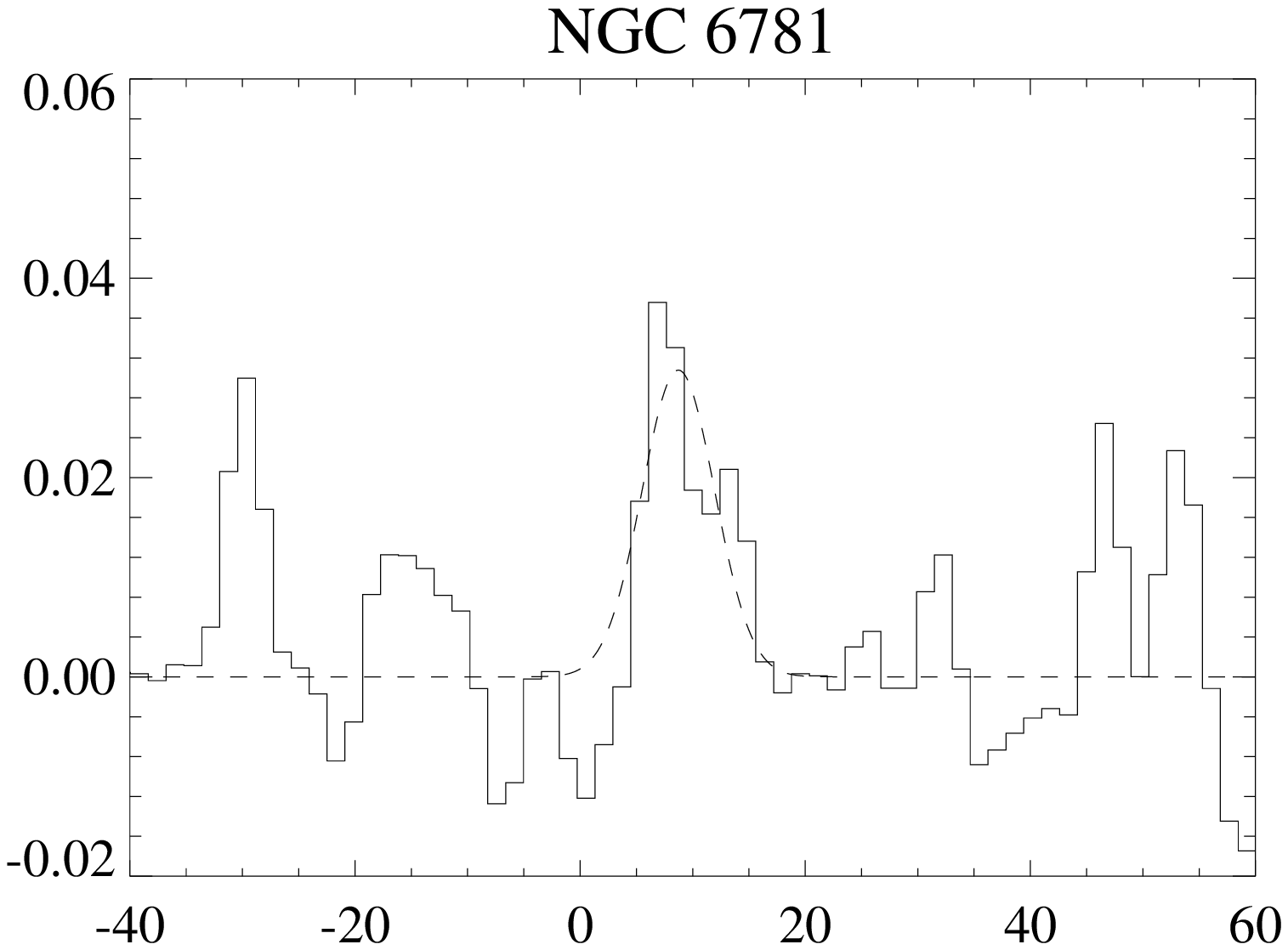}
 \includegraphics[width=0.487\textwidth]{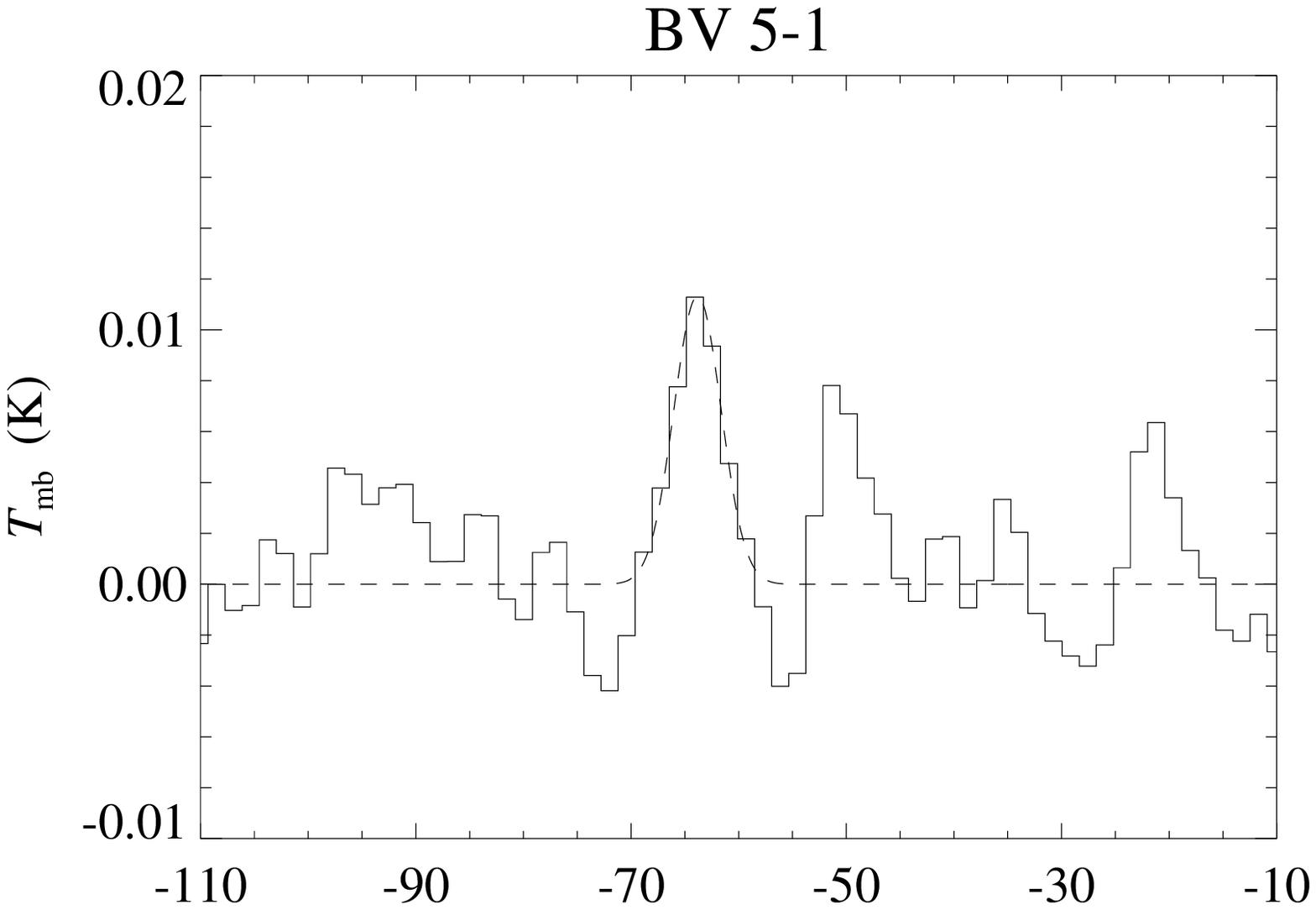}\includegraphics[width=0.463\textwidth]{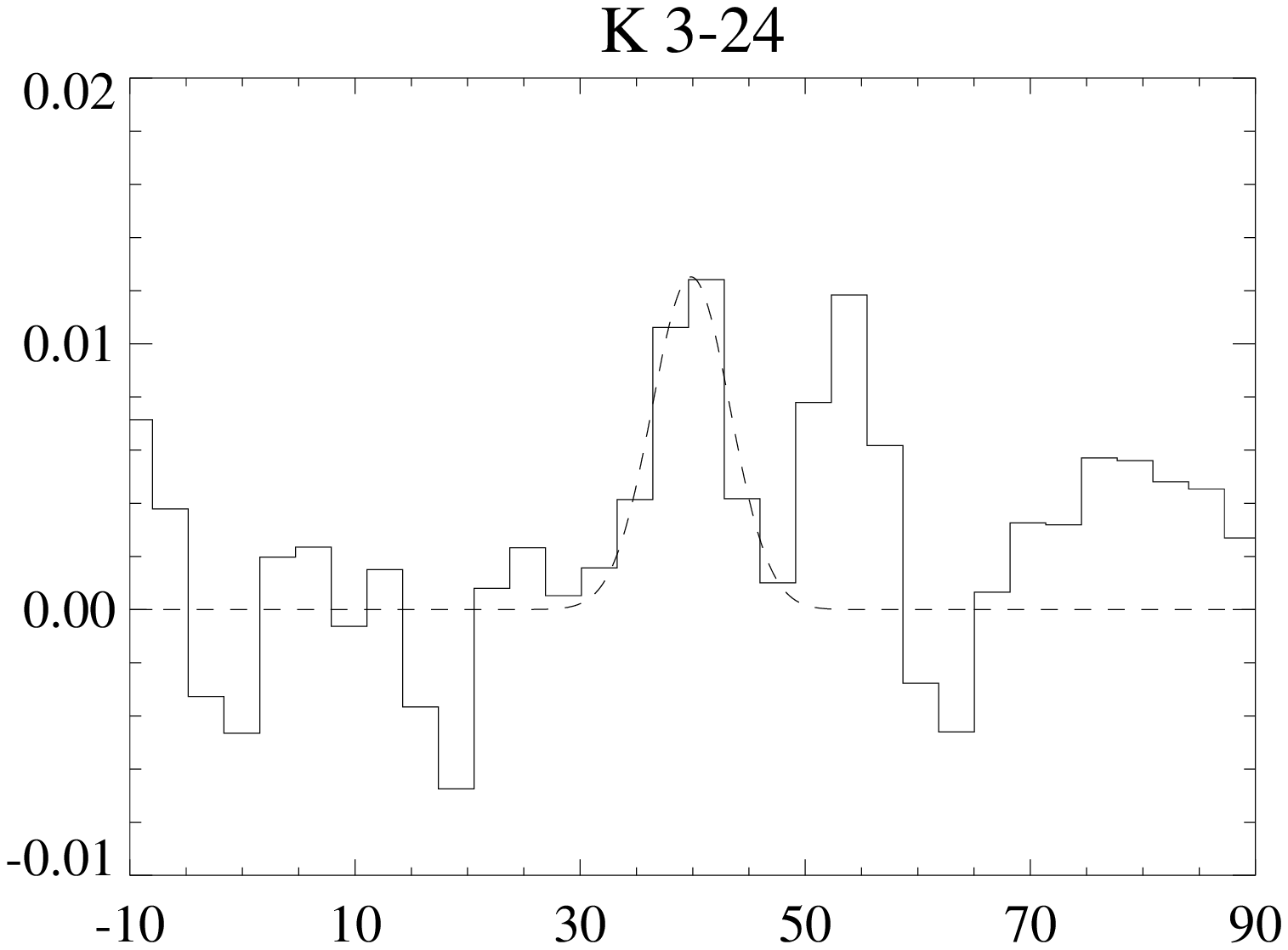}
 \includegraphics[width=0.487\textwidth]{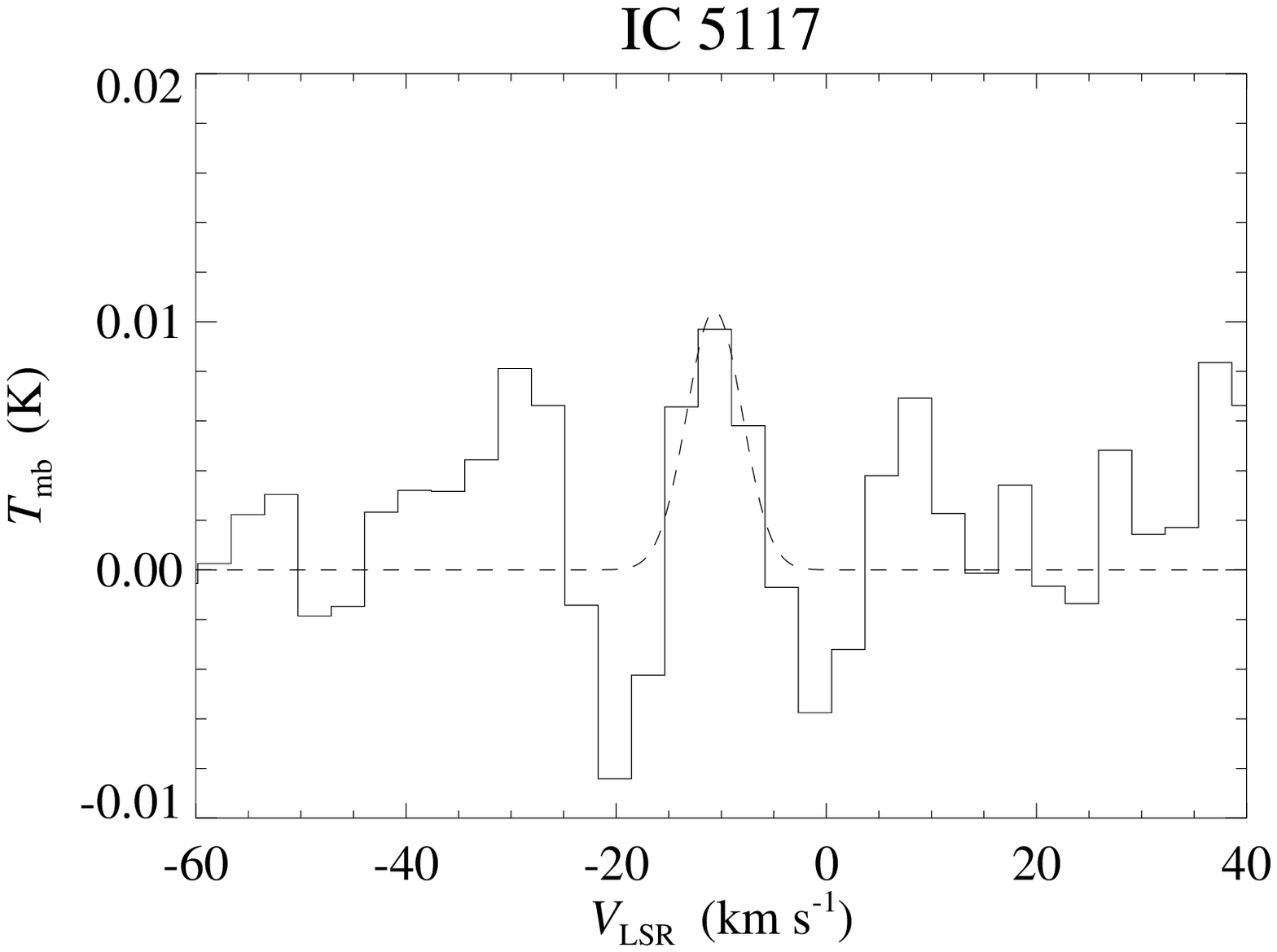}\includegraphics[width=0.463\textwidth]{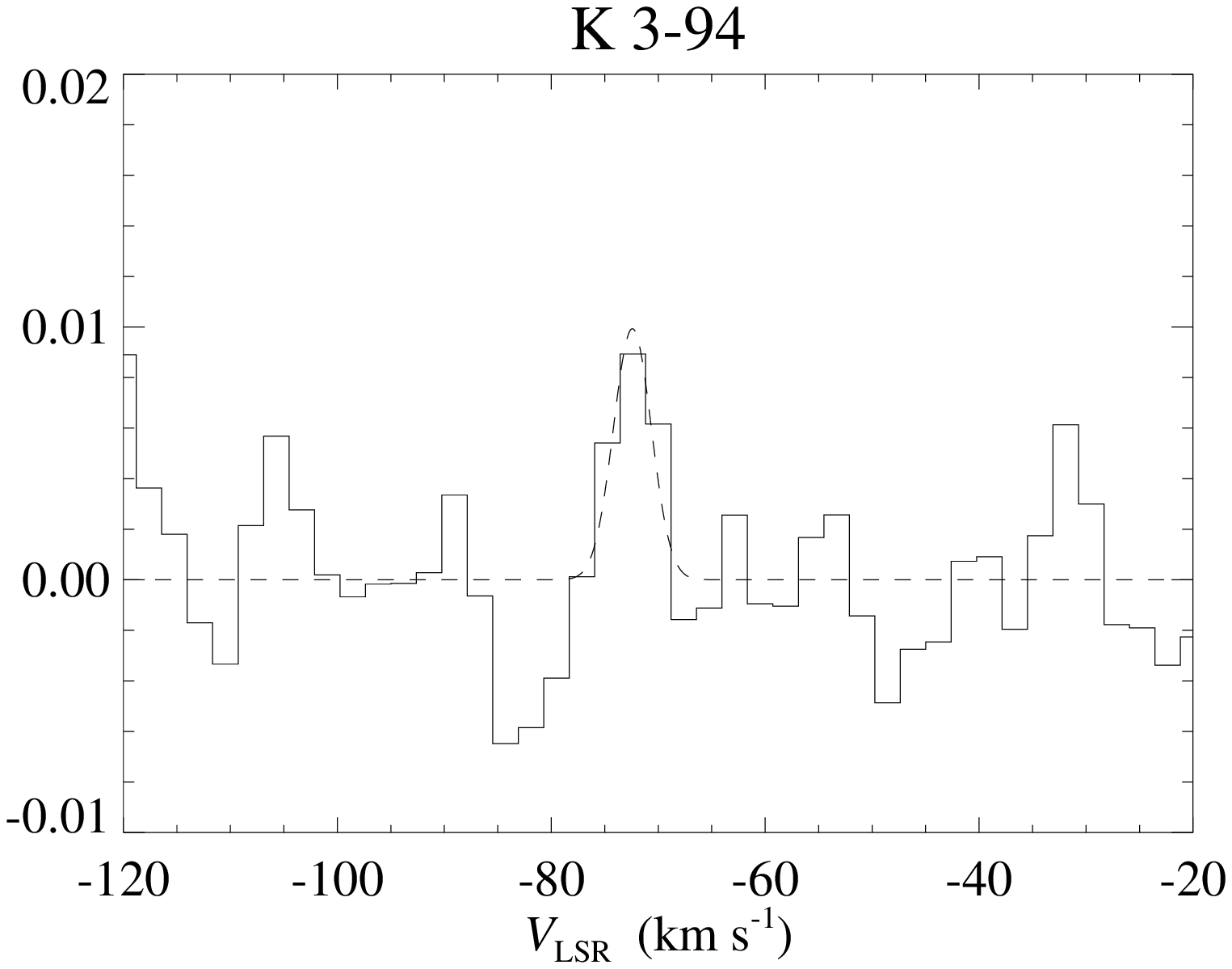}
 \caption{Spectra of the CO$^+$ $N=2\to1$, $J=5/2\to3/2$ line observed in the sample of PNe, with fits to the line profiles overlaid (dashed lines). Brightness temperatures are given in main beam ($T_{\rmn{mb}}$) scale and velocities are quoted with respect to the local standard of rest (LSR). Some of the spectra have been binned to a lower velocity resolution to reduce the noise.}
 \label{spectra1}
\end{figure*}

\begin{table*}
 \caption{Summary of the Gaussian fit parameters for the observed CO$^+$ $N=2\to1$, $J=5/2\to3/2$ line in each source. The integrated intensity, peak main beam temperature, RMS noise, line width, central velocity and the spectral resolution of the final data are listed for each source. In the case of non-detections, an upper limit to the integrated intensity is derived using equation~\ref{upperlimit}.}
 \label{data}
 \begin{tabular}{@{}l c c c c c c c@{\ }}
  \hline
  Source & $\int T_{\rmn{mb}}\,\rmn{d}V$ & $T_{\rmn{mb}}$ & $\sigma_{\rmn{mb}}$ & FWHM & $V_{\rmn{LSR}}$ & $\Delta\nu$ & $\Delta\nu$ \\
  & ($\rmn{K\,km\,s^{-1}}$) & (K) & (K) & ($\rmn{km\,s^{-1}}$) & ($\rmn{km\,s^{-1}}$) & (MHz) & ($\rmn{km\,s^{-1}}$) \\
  \hline
  CRL\,618\textit{a} & 0.762$\pm$0.077 & 0.069 & 0.008 & 10.43 & $-66.03$ & 1.25 & 1.59 \\
  CRL\,618\textit{b} & 0.872$\pm$0.085 & 0.062 & 0.008 & 13.24 & $-24.81$ & 1.25 & 1.59 \\
  NGC\,6781 & 0.249$\pm$0.057 & 0.031 & 0.008 & 7.60 & $\phantom{1}$$+8.73$ & 1.25 & 1.59 \\
  BV\,5--1  & 0.062$\pm$0.015 & 0.011 & 0.003 & 5.18 & $-63.82$ & 1.25 & 1.59 \\ 
  K\,3--24  & 0.109$\pm$0.036 & 0.013 & 0.003 & 8.16 & $+39.86$ & 2.50 & 3.17 \\
  IC\,5117  & 0.069$\pm$0.035 & 0.010 & 0.004 & 6.22 & $-10.64$ & 2.50 & 3.17 \\
  K\,3--94  & 0.044$\pm$0.028 & 0.010 & 0.003 & 4.17 & $-72.41$ & 1.88 & 2.38 \\
  M\,1--17  &        $<$0.351 & 0.006 & 0.003 & {--}-- &\ {--}-- & 2.50 & 3.17 \\
  M\,1--13  &        $<$0.162 & 0.005 & 0.003 & {--}-- &\ {--}-- & 2.50 & 3.17 \\
  \hline
 \end{tabular}
\end{table*}

\begin{figure}
 \includegraphics[width=84mm]{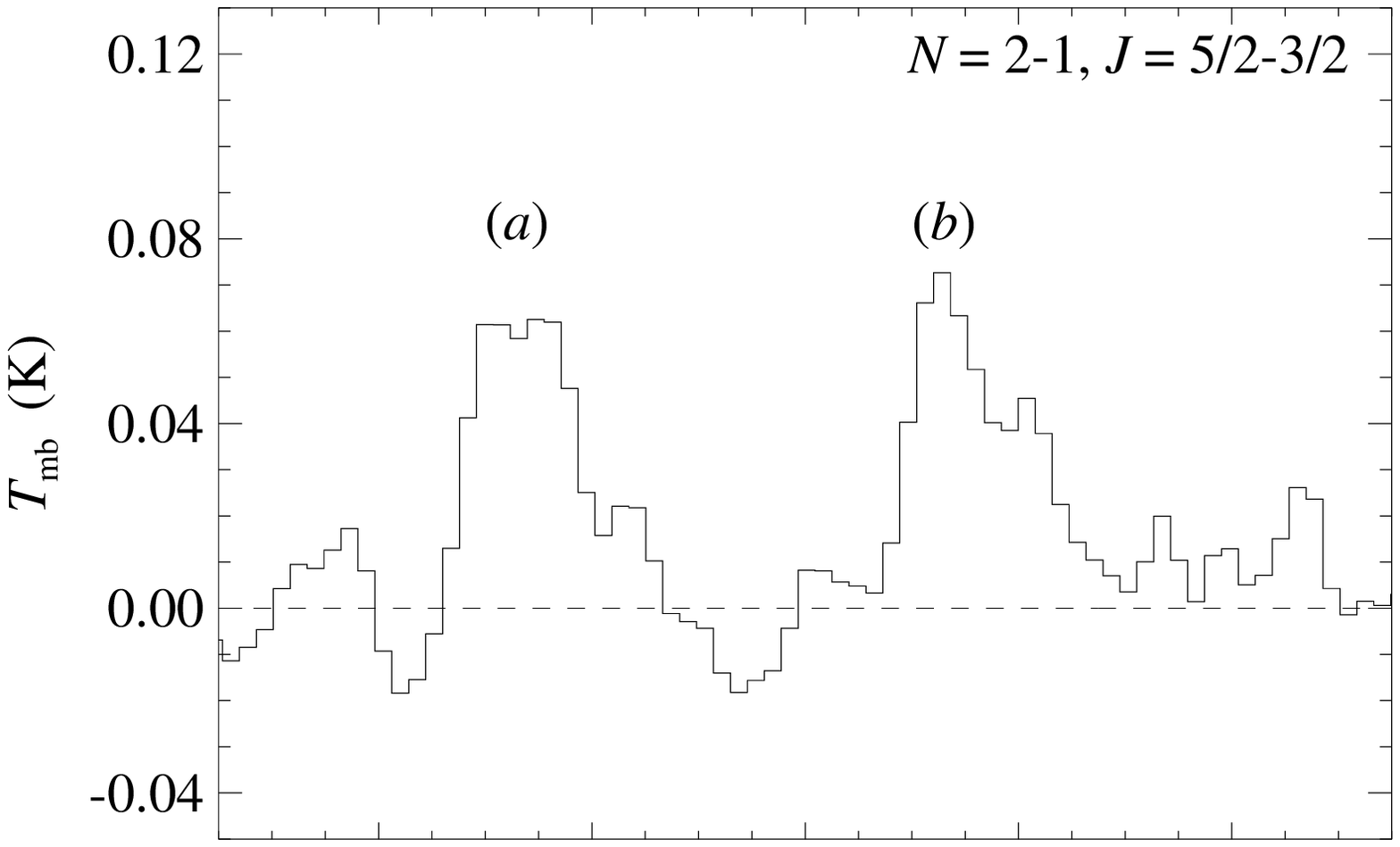}
 \includegraphics[width=84mm]{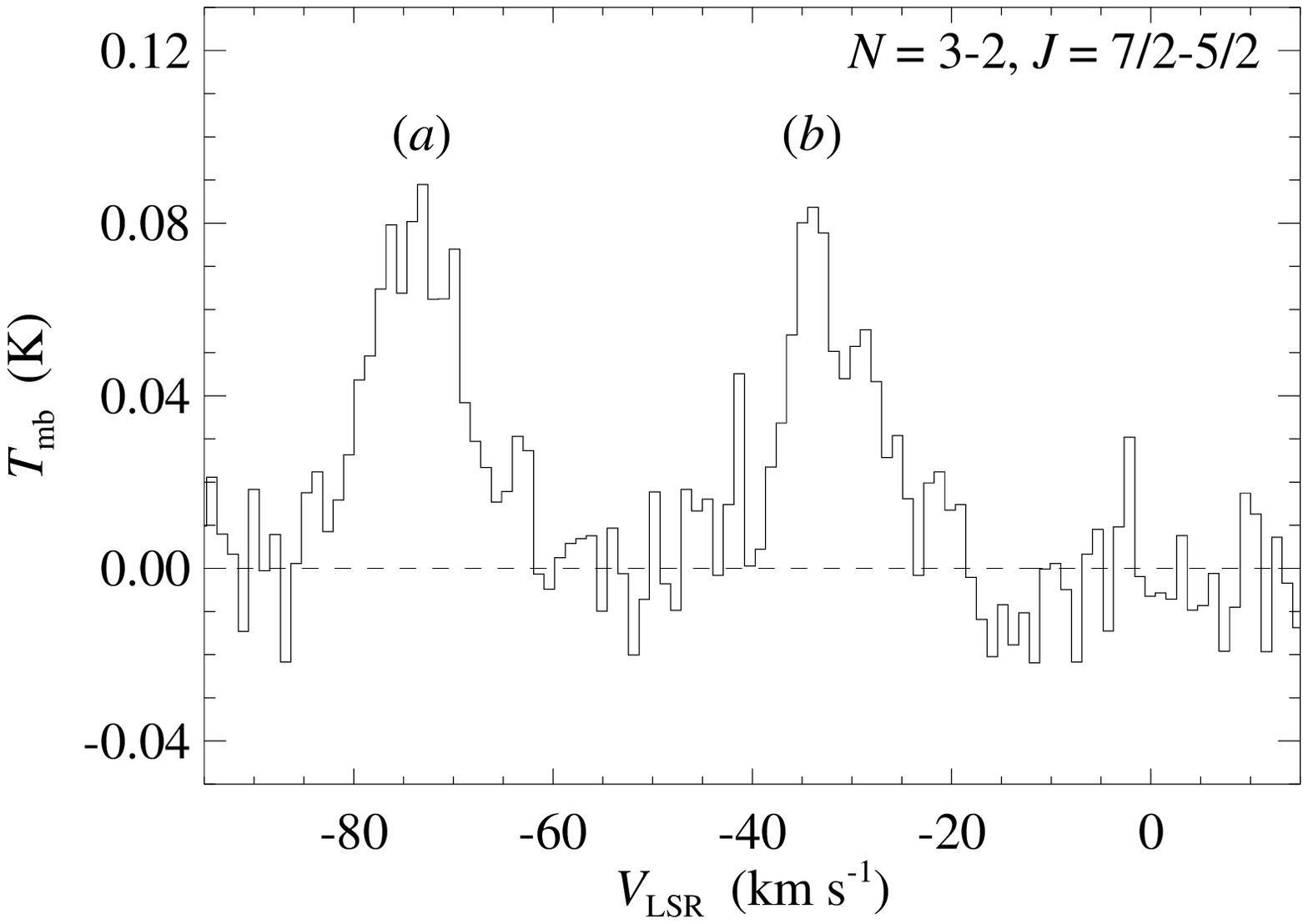}
 \caption{Spectra of the CO$^+$ $N=2\to1$, $J=5/2\to3/2$ and $N=3\to2$, $J=7/2\to5/2$ lines observed in CRL\,618. Brightness temperatures are given in main beam ($T_{\rmn{mb}}$) scale and velocities are quoted with respect to the local standard of rest (LSR). The two velocity components appearing in each of the lines are labelled \textit{a} and \textit{b} on the spectra.}
 \label{spectra2}
\end{figure}

Spectra obtained for the sample of PNe are shown in Figs~\ref{spectra1} and \ref{spectra2}. We have fitted Gaussian profiles to the observed lines and these are overlaid on the spectra in Fig.~\ref{spectra1}. Table~\ref{data} lists the fit parameters for each source, including the integrated intensity (column 2), the peak main beam temperature (3), the RMS noise (4), the line width and central velocity (5, 6), and the spectral resolution of the data (7, 8). For non-detections, upper limits for the line intensities are given by
\begin{equation}\label{upperlimit}
 I(\rmn{CO^{+}}) < 3\sigma_{\rmn{mb}}\Delta V ,
\end{equation}
where $\sigma_{\rmn{mb}}$ is the main beam RMS noise of the spectrum and $\Delta V$ is the assumed line width, taken to be the expansion velocity of the source measured in the CO(2--1) line (see Table~\ref{sources}).

We obtain a strong ($>$\,$10\sigma$) detection of the CO$^+$ $N=2\to1$, $J=5/2\to3/2$ line in CRL\,618 and marginal detections ($\ge$\,$3\sigma$) in another three PNe (NGC\,6781, BV\,5--1, K\,3--24). We obtain tentative detections in IC\,5117 and K\,3--94, and fail to detect the line in the remaining two PNe, M\,1--13 and M\,1--17, for which we derive upper limits to the line intensities. The linewidths of the previously detected HCO$^{+}$ lines in these sources are not given in the literature, but are generally much wider than the CO$^{+}$ linewidths we find here.

Estimates of the column densities have been derived under the assumptions of local thermodynamic equilibrium (LTE) and optically thin emission, following the method outlined in \citet{Fuente03},
\begin{equation}\label{coldenseqn}
 N = 1.94 \times 10^{3} \frac{\nu^{2} Q(T_{\rmn{ex}})}{g_{u}A_{ij}}
 \exp\left(\frac{E_{u}}{kT_{\rmn{ex}}}\right) \int T_{\rmn{mb}}\,\rmn{d}V \times f_{\rmn{b}}^{-1} ,
\end{equation}
where $\int T_{\rmn{mb}}\,\rmn{d}V$ is the integrated intensity of the CO$^+$ line ($\rmn{K\,km\,s^{-1}}$ in main beam scale), $f_{\rmn{b}}$ is the beam filling factor (described below), $\nu$ is the frequency of the transition (in GHz), $Q(T_{\rmn{ex}})$ is the partition function for CO$^+$ at the excitation temperature $T_{\rmn{ex}}$, evaluated by interpolating between values obtained from the Jet Propulsion Laboratory (JPL) molecular line data base \citep{Pickett98}, $E_{u}$ is the upper state energy, $g_{u}$ is the upper state degeneracy and $A_{ij}$ is the Einstein spontaneous emission coefficient for the transition (in s$^{-1}$). The relevant values for the $N=2\to1$, $J=5/2\to3/2$ line are listed in Table~\ref{constants}. The beam filling factor is given by
\begin{equation}\label{fillingfactor}
 f_{\rmn{b}} = \frac{\theta_{\rmn{s}}^{2}}{\theta_{\rmn{s}}^{2} + \theta_{\rmn{b}}^{2}} ,
\end{equation}
where $\theta_{\rmn{s}}$ is the source size (FHWM) and $\theta_{\rmn{b}}$ is the telescope beamwidth (HPBW). For the source sizes, we have adopted the CO(2--1) measurements listed in Table~\ref{sources}. We note that the CO$^{+}$ emission may arise from smaller regions of the nebulae than are traced by CO, however, there are no maps of CO$^{+}$ (and very few of HCO$^{+}$) that may be used, whereas maps of CO emission exist for most of the sources. When sources have not been resolved in previous observations, we have used the upper limits on their size.

Since we would like to correlate the CO$^+$ emission with that of HCO$^+$, we have assumed that the two species are emitted from the same region and we have adopted the same uniform excitation temperature $T_{\rmn{ex}}=25~\rmn{K}$ assumed by \citetalias{Josselin03} in their calculation of the column densities of HCO$^+$, with the exception of CRL\,618, for which we detected two CO$^+$ transitions and can therefore derive an estimate for the excitation temperature using the rotation diagram method (see Sect.~\ref{CRL618}). The chosen temperature of 25~K has been deduced from CO observations in several nebulae \citep{Bachiller93}. As noted before, however, CO is probably ubiquitously present in the molecular gas surrounding these objects, whereas HCO$^+$ and CO$^+$ may only exist in the thin layers of photodissociation regions (PDRs), hence we may be underestimating the kinetic temperature. When the line was not detected, upper limits determined for the integrated intensities were used to derive upper limits for the column densities. The column densities are reported in Table~\ref{columns} where, for reference, we also list those derived previously for HCO$^+$, again corrected for the filling factor (assuming the same source sizes and an IRAM 30-m beamwidth of 28~arcsec). It should be noted that the uncertainties quoted in the table are likely to be overshadowed by larger systematic errors, including uncertainties in the beam filling factors, and may therefore be misleading when comparing the column densities.

\subsection{CRL\,618}\label{CRL618}

The PPN CRL\,618 displays the strongest emission in our sample and has been clearly detected in two lines of CO$^+$ ($N=2\to1$, $J=5/2\to3/2$ and $N=3\to2$, $J=7/2\to5/2$; see Fig.~\ref{spectra2}). This multiple line detection allows the rotational temperature of the gas to be determined with a rotational (Boltzmann) diagram using the method detailed by \citet{Fuente03}. A straight line fit to the rotational diagram yields a rotational temperature of 32~K. This higher temperature is similar to the assumed excitation temperature of 25~K \citep{Bachiller93}, but lower than would be expected for a PDR region. If taken to be representative of the temperature for all of our sample, an excitation temperature of 32~K reduces the column densities presented in Table~\ref{columns} by 14 per cent.

Two velocity components are clearly seen at $-70$ and $-30~\rmn{km\,s^{-1}}$ in Fig.~\ref{spectra2}, in both transitions of CO$^+$. These velocities are consistent with the interferometric observations of HCO$^+$ and HCN by \citet{Sanchez04a}. They detected HCO$^+$ $J=1\to0$ and HCN $J=1\to0$ emission across a velocity range $-120 \le V_{\rmn{LSR}} \le +40~\rmn{km\,s^{-1}}$, arising from two distinct regions, which they assigned to a high-velocity bipolar outflow and a slowly expanding equatorial torus. We identify the two blue-shifted features in the CO$^+$ transitions as follows. Feature \textit{a} is the fast component, arising in the blue-shifted (i.e. brightest) lobe of the bipolar outflow, and feature \textit{b} is due to the slowly expanding torus. We propose that the red-shifted lobe of the bipolar outflow is not visible in CO$^+$ emission because it is much fainter, as is the case for the HCO$^+$ emission observed by \citet{Sanchez04a}. Their fluxes were 2.7 (red) versus 6.4 (blue) Jy\,beam$^{-1}$\,km\,s$^{-1}$. The slight offset in peak velocities in the two transitions in Fig.~\ref{spectra2} is likely due to differing excitation conditions in the gas; the HCO$^+$ and HCN lines in \citet{Sanchez04a} also peak at slightly different velocities. Interestingly, our CO$^+$ line components are of comparable strengths. This contrasts with the HCO$^+$ and HCN lines, in which the outflow component is much weaker than the torus component \citep[][their figure 1]{Sanchez04a}, and may indicate that different CO$^+$ formation and destruction routes operate in these two regions.

\begin{table}
 \begin{center}
  \begin{minipage}{63mm}
   \caption{Molecular data for CO$^+$.}
   \label{constants}
   \begin{tabular}{@{}lcc}
    \hline
    Parameter        & $N=2\to1$,    & $N=3\to2$,    \\
                     & $J=5/2\to3/2$ & $J=7/2\to5/2$ \\
    \hline
    $\nu$~(GHz)      & 236.0626      & 354.0142 \\
    $E_{u}$~(K)      & 16.997        & 33.987   \\
    $\mu$~(Debye)    & 2.771         & 2.771    \\
    $Q(25\,\rmn{K})$ & 18.35         & 18.35    \\
    $g_{u}$          & 6             & 8        \\
    $A_{ij}$~(s$^{-1}$) & 4.7$\times$10$^{-4}$ & 1.7$\times$10$^{-3}$ \\
    \hline
   \end{tabular}
  \end{minipage}
 \end{center}
\end{table}

\begin{table}
 \begin{center}
  \begin{minipage}{63mm}
   \caption{Estimated column densities of CO$^+$ and HCO$^+$.}
   \label{columns}
   \begin{tabular}{@{}l r@{$\times$}l r@{$\times$}l}
    \hline
    Source & \multicolumn{2}{c}{$N(\rmn{CO^+})$}   & \multicolumn{2}{c}{$N(\rmn{HCO^+})$}  \\
           & \multicolumn{2}{c}{($\rmn{cm^{-2}}$)} & \multicolumn{2}{c}{($\rmn{cm^{-2}}$)} \\
    \hline
    CRL\,618  & 3.7$\pm$0.4 & 10$^{12}$ & 1.6$\pm$0.1 & 10$^{14}$$^{\star}$ \\
    NGC\,7027 & 1.8$\pm$0.4 & 10$^{12}$$^{\ddag}$ & 4.3$\pm$0.1 & 10$^{13}$$^{\dag}$ \\
    M\,1--17  &      $<$1.8 & 10$^{12}$ & 1.1$\pm$0.1 & 10$^{13}$$^{\dag}$  \\
    IC\,5117  & 3.6$\pm$1.8 & 10$^{11}$ & 3.5$\pm$0.6 & 10$^{12}$$^{\dag}$  \\
    M\,1--13  &      $<$8.5 & 10$^{11}$ & 9.0$\pm$0.6 & 10$^{12}$$^{\dag}$  \\
    BV\,5--1  & 3.7$\pm$0.9 & 10$^{11}$ & 1.3$\pm$0.2 & 10$^{12}$$^{\dag}$  \\
    K\,3--94  & 3.1$\pm$2.0 & 10$^{11}$ & 2.3$\pm$0.3 & 10$^{12}$$^{\dag}$  \\
    K\,3--24  & 7.6$\pm$2.5 & 10$^{11}$ & 5.3$\pm$0.3 & 10$^{12}$$^{\dag}$  \\
    NGC\,6781 & 3.5$\pm$0.8 & 10$^{11}$ & 5.5$\pm$0.5 & 10$^{12}$$^{\star}$ \\
    \hline
   \end{tabular}

   \medskip
   $^{\star}$Values taken from \citet{Bachiller97}\\
   $^{\dag}$Values taken from \citet{Josselin03}\\
   $^{\ddag}$Calculated using the CO$^+$ integrated intensity observed by \citet{Latter93}
  \end{minipage}
 \end{center}
\end{table}


\section{HCO$\bmath{^+}$ as an evolutionary tracer?}\label{Discussion}

\begin{figure}
 \includegraphics[angle=90, width=84mm]{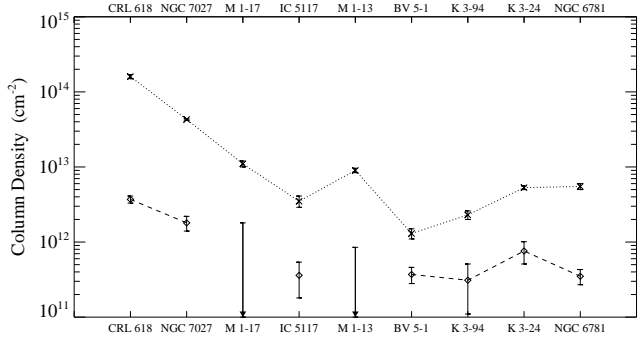}
 \caption{Column densities (with error-bars) of CO$^+$ (diamonds) and HCO$^+$ (crosses, from the literature) for the sample of planetary nebulae. The sources span a range of evolutionary stages from PPN (left) through to evolved PN (right). Only an upper limit was obtained for the column density of CO$^+$ for M\,1--17 and M\,1--13.}
 \label{comparison}
\end{figure}

\begin{figure}
 \includegraphics[angle=90, width=84mm]{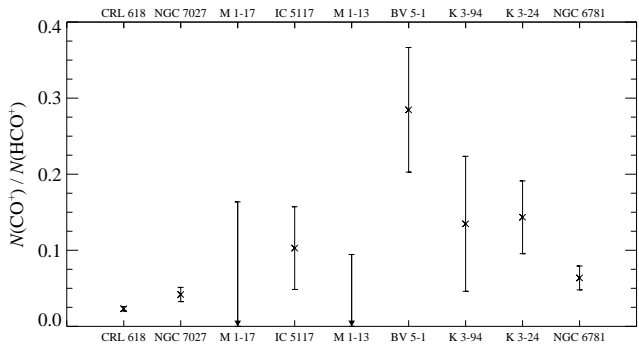}
 \caption{Ratios of CO$^{+}$ to HCO$^{+}$ column density (with error-bars) for the sample of planetary nebulae. The sources span a range of evolutionary stages from PPN (left) through to evolved PN (right). Only an upper limit was obtained for the column density of CO$^+$ for M\,1--17 and M\,1--13.}
 \label{ratios}
\end{figure}

It is impossible at this stage to establish if and when CO$^+$ is the source of the HCO$^+$ enhancement. Nevertheless, despite the lack of a statistically meaningful sample, in Figs~\ref{comparison} and \ref{ratios} we plot the derived column densities of CO$^+$ and HCO$^+$ (the latter taken from the literature) for our sample of planetary nebulae, including NGC\,7027 (for which we take the CO$^+$ data from \citealt{Latter93} and HCO$^+$ data from \citetalias{Josselin03}) as a function of their believed evolutionary stage, from PPNe (left) to evolved PNe (right), as well as their ratios. The evolutionary stage of each PNe is estimated based on the ratio of the molecular mass to the ionized mass $M_{\rmn{m}}/M_{\rmn{i}}$, which is believed to be a good indicator and almost independent of distance \citep{Huggins96}. Fig.~\ref{comparison} does seem to indicate a correlation between HCO$^+$ and CO$^+$. For example, both CRL\,618 and NGC\,7027, believed to be young objects, have the highest HCO$^+$ and the highest CO$^+$ abundances (in fact they are the only 2 objects for which a strong detection of CO$^+$ has been reported). This implies that, although young and hence enough H$_3^+$ should still be available to form HCO$^+$, CO$^+$ is indeed available and, at early evolutionary stages, must be formed by charge transfer between H$^+$ and CO. It is then also interesting to note that as we move along the evolutionary $x$-axis, after a dip, both HCO$^+$ and CO$^+$ tend to increase with age. This may mean that at later stages, when the propagation front appears, more CO$^+$ (and hence HCO$^+$) is formed (via the second route of formation). We cannot draw any conclusions from the 2 objects for which CO$^+$ was not detected (for which we derive an upper limit) but for which HCO$^+$ is quite high. A possible explanation for the lack of CO$^+$ is that at this evolutionary stage the radiation is strong enough to destroy this species, whilst there is still enough H$_3^+$ to make HCO$^+$.

\section{Conclusions}\label{Conclusions}

We have searched for the molecular ion CO$^+$ in a sample of PPNe and PNe where HCO$^+$ was previously detected. If detected, CO$^+$ may help our understanding of the origin of the strong HCO$^+$ emission. We obtain a strong detection of CO$^+$ in two transitions in CRL\,618 and $\ge$\,$3\sigma$ detections of the lower transition in three other objects, namely NGC\,6781, BV\,5--1 and K\,3--24. The lower CO$^+$ transition is also weakly detected in IC\,5117 and K\,3--94, but is not detected in M\,1--13 and M\,1--17. We find that the integrated intensity of the CO$^+$ line is reasonably well correlated with that of HCO$^+$. Although we cannot draw any strong conclusion on the origin of the HCO$^+$ emission, we can conclude that CO$^+$ is present in at least some early PNe and hence this may provide a supplementary source for HCO$^+$ and that, in evolved PNe, where H$_3^+$ should be almost exhausted, CO$^+$ may be present in high enough abundances to boost HCO$^+$ again.


\section*{Acknowledgements}
We thank the referee for constructive comments which helped to improve an earlier draft of this paper. The James Clerk Maxwell Telescope is operated by The Joint Astronomy Centre on behalf of the Science and Technology Facilities Council of the United Kingdom, the Netherlands Organisation for Scientific Research, and the National Research Council of Canada. The data were obtained under the programs M05AU10 and M05BU78. TAB and WW are supported by PPARC studentships. SV acknowledges individual financial support from a PPARC Advanced Fellowship.


\bsp

\label{lastpage}

\end{document}